\documentclass[10pt,conference,twocolumn]{IEEEtran}

\usepackage{latexcad}
\usepackage{amsmath}
\usepackage{amssymb}
\usepackage{latexsym}
\usepackage{amsfonts}

\begin{document}

\title{Embedding Constructions of Tail-Biting Trellises for Linear Block Codes}

\author{
\authorblockN{ Jianqin Zhou}
\authorblockA{1. Telecommunication School, Hangzhou Dianzi University,
Hangzhou, 310018 China\\
2. Department of Computer Science, Anhui Univ. of Technology,Ma'anshan, 243002 China\\
zhou9@yahoo.com
 }
 }

\maketitle

\begin{abstract}

In this paper, embedding construction of tail-biting trellises for
linear block codes is presented. With the new approach of
constructing  tail-biting trellises, most of the study of
tail-biting trellises can be converted into the study of
conventional trellises. It is proved that any minimal tail-biting
trellis can be constructed by the recursive process of embedding
constructions from the well-known Bahl-Cocke-Jelinek-Raviv (BCJR)
constructed conventional trellises. Furthermore, several properties
of embedding constructions of tail-biting trellises are discussed.
Finally, we give four sufficient conditions to reduce the maximum
state-complexity of a trellis with one peak.\\

\end{abstract}

\section{Introduction}

To reduce decoding complexity of a linear block code, in the papers
\cite{BCJR,Wolf,Forney,Kschischang95,Kschischang96,Muder} and
references therein, conventional trellis representations of a linear
block code have been proposed and investigated extensively. With
these representations, different efficient soft-decision decodings
of codes can be applied to decode a linear block code, for example,
the Viterbi algorithm.

To further reduce the complexity, just as indicated and studied in
the papers \cite{Koetter02,Koetter03,Kschischang96,McEliece,Muder},
characterizing and constructing minimal trellises for conventional
trellis representations have key of importance. Based on this
consideration, tail-biting trellises for a linear block code have
been appeared. Although much unknown for these trellis still remain,
the papers \cite{Calderbank}, \cite{Wiberg} have shown that the
number of states in a tail-biting trellis for a linear code can be
as low as the square root of the number of states which is used in
the minimal conventional trellis. These results have greatly
activated the interests and concerns of many researchers. In recent
years, much advance has been made in this direction, for example,
see \cite{Koetter02,Koetter03,Shankar01,Shankar03,Nori,Shany} and
the references therein.

Differing to a conventional trellis representation of a linear block
code, a tail-biting trellis representation may have several starting
and ending status pairs, which helps to reduce the total status
number and hence, reduce the decoding complexity, while there is
only one starting and ending status pair in a conventional
representation. Just because of this, there have more flexible
designs of tail-biting trellis representations, and at the same
time, it is more difficult to find out the optimal representation
for any linear block code. Here, the optimality means that there is
the smallest status in the trellis. In fact, a method to design the
optimal trellis for any linear block code has not appeared until
now. Fortunately, there are a lot of works on this direction.
Koetter and Vardy, in the papers \cite{Koetter02}, \cite{Koetter03},
have made a detailed study of the structure of linear tail-biting
trellises. In the paper \cite{Nori}, the authors followed the idea
given in the papers \cite{BCJR}, \cite{Forney}, presented new ways
of describing and constructing linear tail-biting trellises for
linear block codes. By following their consideration, the minimal
tail-biting trellis computation problem may thus be formulated as
the problem to find a suitable matrix. However, to find this
suitable matrix still is a difficult task, moreover, the paper did
not give any method to overcome this difficulty.

In this paper, we will demonstrate that an embedding construction of
a tail-biting trellis can be converted into a construction of a
conventional trellis. It turns out that many properties of a
conventional trellis can be switched into ones of a tail-biting
trellis. Thus, a tail-biting trellis can be obtained by using a
corresponding conventional trellis for a given linear block code.
Furthermore, we will prove that any minimal tail-biting trellis can
be constructed by the recursive process of embedding constructions
from the well-known BCJR constructed conventional trellises. Based
on the conclusions above, moreover, several properties of embedding
constructions of tail-biting trellises are discussed in this paper.
Finally, we also will give four sufficient conditions to reduce the
maximum state-complexity of a conventional or a tail-biting trellis.

The organization of this paper is as follows. In the next section,
some preliminaries are given, and in the section III, the embedding
method and main results are stated. Four sufficient conditions are
presented in the section IV. At last, conclusions are given in the
section V.

%

\section{Preliminaries}

In this section, a number of definitions and concepts related to
conventional and tail-biting trellises will be introduced. We will
follow some notations and definitions in \cite{Koetter03} and
\cite{Nori}.

Firstly, we need a few of terminologies from graph theory. An
edge-labeled directed graph is defined as a triple $(V,E,\Sigma)$,
which consists of a set $V$ of vertices, a finite set $\Sigma$, and
a set $E$ of ordered triples $(u, a, v)$, with $u, v\in V$ and
$a\in\Sigma$. Usually, $\Sigma$ is called as the alphabet and $(u,
a, v)$ is called as an edge. Also an edge $(u, a, v)\in E$ means
that it begins at $u$, ends at $v$, and has label $a$.

The following definitions are also necessary for this paper.

\noindent {\bf Definition 1}: A conventional trellis
$T=(V,E,\Sigma)$ of depth $n$ is an edge-labeled directed graph,
which satisfies the following property: the set $V$ can be
partitioned into $n+1$ vertex classes, denoted as
\begin{equation}\label{formula001}
V=V_0\cup V_1\cup\cdots\cup V_n,
\end{equation}
where $|V_0| = |V_n| = 1$, such that every edge in $E$ is labeled
with a symbol from the alphabet $\Sigma$, and begins at a vertex of
$V_i$ and ends at a vertex of $V_{i+1}$, for some
$i\in\{0,1,\ldots,n-1\}$. The ordered index set $I = \{0, 1, \ldots
, n\}$ introduced by the partition of $V$ in (\ref{formula001}) is
called the time indices for $T$.

A conventional trellis $T$ is reduced if every vertex in $T$ lies on
at least one path from a vertex in $V_0$ to a vertex in $V_n$.



\noindent {\bf Definition 2}: A tail-biting trellis $T=(V,E,\Sigma)$
of depth $n$ is an edge-labeled directed graph, if it satisfies
condition that the set $V$ can be partitioned into $n$ vertex
classes
\begin{eqnarray}
V=V_0\cup V_1\cup\cdots\cup V_{n-1} \label{formula002},
\end{eqnarray}
such that every edge in $T$ is labeled with a symbol from the
alphabet $\Sigma$, and begins at a vertex of $V_i$ and ends at a
vertex of $V_{i+1(\mbox{ mod }n)}$, for some
$i\in\{0,1,\ldots,n-1\}$.

Some remarks are required here. The first, from the definitions, it
is obvious that a conventional trellis is a tail-biting trellis, but
the inverse is not true. The second, in a conventional trellis, the
sizes of $V_{0}$ and $V_{n}$ are all equal to $1$. In contrast to
this, there is no such requirement in a tail-biting trellis.
Moreover, if the size of $V_{0}$ is equal to $1$, a tail-biting
trellis is reduced to a conventional one. The third, if an edge
begins at a vertex in $V_{n-1}$, it will end at a vertex in $V_{0}$
in a tail-biting trellis, on the contrast, it will end at a vertex
in $V_{n}$ in a conventional one.

We continue to define some terminologies. The indices in the set
$I=\{0,1,\ldots,n-1\}$ for the partition in (\ref{formula002}) are
called as the time indices. Moreover, in this paper, the set $I$ is
identified with $\mathbb{Z}_n$, the residue classes of integers
modulo $n$. And hence, an interval of indices [$i,j$] means the
sequence $\{i,i+1,\ldots,j \}$ if $i<j$, and the
sequence$\{i,i+1,\ldots,n-1,0,\ldots, j \}$ if $i>j$. Every cycle of
length $n$ in $T$ starting at a vertex of $V_0$ defines a vector
$(a_0,a_1,\ldots,a_{n-1})\in \Sigma^n$, which is an edge-label
sequence. If every vertex in $T$ lies on at least one cycle from a
vertex in $V_0$, the tail-biting trellis $T$ is defined as reduced.

Secondly, some connections between a linear block code and an
edge-labeled directed graph are needed. According the results given
in papers \cite{BCJR,Forney,Kschischang95,Muder,McEliece,Wolf}, any
linear block code can be represented by using a conventional trellis
or a tail-biting trellis. Let us make these representations more
precisely.

Denote an $(n,k)$ linear block code over $F_q$ as $(n,k)_q$. Assume
that $C=(n,k)_{q}$ is a linear block code. Thus, every codeword in
$C$ is a vector over $F_{q}$ with size $n$. Arranging all entries in
this vector in the natural order becomes a sequences in $F_{q}$ with
length $n$. If the set consisting of all these sequences is
precisely the same as the one consisting of all edge-labeled
sequences corresponding to those cycles in $T$ that start at a
vertex of $V_0$, the conventional or tail-biting trellis $T$ is said
to represent a block code $C$ of length $n$ over $\Sigma (=F_{q})$.

Recall the facts that the number of states in a trellis code is an
important factor in Viterbi decoding and it is directly related to
decoding complexity. Hence, the quantity $\log_{|\Sigma|}|V_i|$ is
regarded as the state-complexity of the trellis, either conventional
or tail-biting, at time index $i$. At the same time, the sequence
$\{\log_{|\Sigma|}|V_i|, 0\le i< n\}$ gives the state-complexity
profile (SCP) of the trellis. Therefore, a trellis $T$ is said to be
minimal if the maximum state-complexity over all time indices
denoted by $s_{\max}(T)$ is minimized over all possible coordinate
permutations of the code \cite{Muder}. In the same paper, it is
proved that the minimal conventional trellis for a linear block code
is unique, and simultaneously, satisfies all definitions of
minimality. Moreover, it is also biproper (that is, any pair of
edges directed towards a vertex has distinct labels, and so also any
pair of edges leaving a vertex).

To help an understanding of the notations and concepts above, the
trellis shown in Fig. 1 is the minimal conventional trellis for the
$(7,4)_2$ Hamming code, which has a parity check matrix defined as
follows:
\begin{eqnarray*}
H= \left[
\begin{array}{ccccccc}
1 & 1&0&0&1&0&1\\
1 & 1&1&0&0&1&0\\
0 & 1&1&1&0&0&1
\end{array}\right].
\end{eqnarray*}


\unitlength=0.11mm

\begin{picture}(500,1000)(20,-530)

\scriptsize

\drawpath{50}{0}{100}{100}\put(70,70){$1$} \put(100,100){$
\begin{array}{c}|1|\\|1|\\|0|\end{array}$}

\drawpath{150}{100}{200}{100}\put(180,100){$0$}\put(200,100){$
\begin{array}{c}|1|\\|1|\\|0|\end{array}$}

\drawpath{250}{200}{300}{-200}\put(280,0){$0$}\put(300,-200){$
\begin{array}{c}|0|\\|0|\\|1|\end{array}$}

\drawpath{250}{200}{300}{300}\put(295,260){$1$}\put(300,300){$
\begin{array}{c}|0|\\|1|\\|0|\end{array}$}

\drawpath{150}{100}{200}{200}\put(180,150){$1$}\put(200,200){$
\begin{array}{c}|0|\\|0|\\|1|\end{array}$}

\drawpath{250}{100}{300}{-400}\put(280,-390){$1$}\put(300,-400){$
\begin{array}{c}|1|\\|0|\\|1|\end{array}$}

\drawpath{250}{100}{300}{100}\put(280,100){$0$}\put(300,100){$
\begin{array}{c}|1|\\|1|\\|0|\end{array}$}


\drawpath{50}{0}{100}{-100}\put(70,-70){$0$}\put(100,-100){$
\begin{array}{c}|0|\\|0|\\|0|\end{array}$}

\drawpath{150}{-100}{200}{-100}\put(180,-100){$0$}\put(200,-100){$
\begin{array}{c}|0|\\|0|\\|0|\end{array}$}

\drawpath{250}{-100}{300}{-100}\put(280,-100){$0$}\put(300,-100){$
\begin{array}{c}|0|\\|0|\\|0|\end{array}$}

\drawpath{250}{-100}{300}{400}\put(290,400){$1$}\put(300,400){$
\begin{array}{c}|0|\\|1|\\|1|\end{array}$}

\drawpath{150}{-100}{200}{-200}\put(180,-150){$1$}\put(200,-200){$
\begin{array}{c}|1|\\|1|\\|1|\end{array}$}

\drawpath{250}{-200}{300}{-300}\put(260,-250){$1$}\put(300,-300){$
\begin{array}{c}|1|\\|0|\\|0|\end{array}$}

\drawpath{250}{-200}{300}{200}\put(255,-190){$0$}\put(300,200){$
\begin{array}{c}|1|\\|1|\\|1|\end{array}$}



\drawpath{350}{300}{400}{300}\put(380,300){$0$}\put(400,300){$
\begin{array}{c}|0|\\|1|\\|0|\end{array}$}

\drawpath{350}{100}{400}{100}\put(380,100){$0$}\put(400,100){$
\begin{array}{c}|1|\\|1|\\|0|\end{array}$}

\drawpath{350}{200}{400}{100}\put(380,150){$1$}\put(400,100){$
\begin{array}{c}|1|\\|1|\\|0|\end{array}$}

\drawpath{350}{200}{400}{200}\put(380,200){$0$}\put(400,200){$
\begin{array}{c}|1|\\|1|\\|1|\end{array}$}

\drawpath{350}{100}{400}{200}\put(380,150){$1$}\put(400,200){$
\begin{array}{c}|1|\\|1|\\|1|\end{array}$}

\drawpath{350}{400}{400}{300}\put(380,350){$1$}\put(400,300){$
\begin{array}{c}|0|\\|1|\\|0|\end{array}$}

\drawpath{350}{400}{400}{400}\put(380,400){$0$}\put(400,400){$
\begin{array}{c}|0|\\|1|\\|1|\end{array}$}

\drawpath{350}{300}{400}{400}\put(380,350){$1$}\put(400,400){$
\begin{array}{c}|0|\\|1|\\|1|\end{array}$}



\drawpath{350}{-100}{400}{-100}\put(380,-100){$0$}\put(400,-100){$
\begin{array}{c}|0|\\|0|\\|0|\end{array}$}

\drawpath{350}{-200}{400}{-100}\put(380,-150){$1$}\put(400,-100){$
\begin{array}{c}|0|\\|0|\\|0|\end{array}$}

\drawpath{350}{-200}{400}{-200}\put(380,-200){$0$}\put(400,-200){$
\begin{array}{c}|0|\\|0|\\|1|\end{array}$}

\drawpath{350}{-100}{400}{-200}\put(380,-150){$1$}\put(400,-200){$
\begin{array}{c}|0|\\|0|\\|1|\end{array}$}

\drawpath{350}{-300}{400}{-300}\put(380,-300){$0$}\put(400,-300){$
\begin{array}{c}|1|\\|0|\\|0|\end{array}$}

\drawpath{350}{-400}{400}{-300}\put(380,-350){$1$}\put(400,-300){$
\begin{array}{c}|1|\\|0|\\|0|\end{array}$}

\drawpath{350}{-400}{400}{-400}\put(380,-400){$0$}\put(400,-400){$
\begin{array}{c}|1|\\|0|\\|1|\end{array}$}

\drawpath{350}{-300}{400}{-400}\put(380,-350){$1$}\put(400,-400){$
\begin{array}{c}|1|\\|0|\\|1|\end{array}$}


\drawpath{450}{200}{500}{200}\put(480,200){$0$}\put(500,200){$
\begin{array}{c}|1|\\|1|\\|1|\end{array}$}

\drawpath{450}{400}{500}{200}\put(480,300){$1$}\put(500,200){$
\begin{array}{c}|1|\\|1|\\|1|\end{array}$}

\drawpath{450}{-200}{500}{-200}\put(480,-200){$1$}\put(500,-200){$
\begin{array}{c}|1|\\|0|\\|1|\end{array}$}

\drawpath{450}{-400}{500}{-200}\put(480,-300){$0$}\put(500,-200){$
\begin{array}{c}|1|\\|0|\\|1|\end{array}$}

\drawpath{450}{-100}{500}{-100}\put(480,-100){$0$}\put(500,-100){$
\begin{array}{c}|0|\\|0|\\|0|\end{array}$}

\drawpath{450}{-300}{500}{-100}\put(480,-200){$1$}\put(500,-100){$
\begin{array}{c}|0|\\|0|\\|0|\end{array}$}

\drawpath{450}{300}{500}{100}\put(480,200){$0$}\put(500,100){$
\begin{array}{c}|0|\\|1|\\|0|\end{array}$}

\drawpath{450}{100}{500}{100}\put(480,100){$1$}\put(500,100){$
\begin{array}{c}|0|\\|1|\\|0|\end{array}$}


\drawpath{550}{-200}{600}{-100}\put(560,-150){$0$}\put(600,-100){$
\begin{array}{c}|1|\\|0|\\|1|\end{array}$}

\drawpath{550}{200}{600}{-100}\put(580,-50){$1$}\put(600,-100){$
\begin{array}{c}|1|\\|0|\\|1|\end{array}$}

\drawpath{550}{-100}{600}{100}\put(590,80){$0$}\put(600,100){$
\begin{array}{c}|0|\\|0|\\|0|\end{array}$}

\drawpath{550}{100}{600}{100}\put(590,100){$1$}\put(600,100){$
\begin{array}{c}|0|\\|0|\\|0|\end{array}$}


\drawpath{650}{-100}{700}{0}\put(680,-50){$1$}\put(700,00){$
\begin{array}{c}|0|\\|0|\\|0|\end{array}$}

\drawpath{650}{100}{700}{1}\put(680,50){$0$}\put(700,00){$
\begin{array}{c}|0|\\|0|\\|0|\end{array}$}


\put(0,00){$
\begin{array}{c}|0|\\|0|\\|0|\end{array}$}

\put(130,-480){\small Fig.1 The minimal conventional trellis}

\put(200,-505){\small for a $(7,4)_2$ Hamming code.}

\end{picture}

From the figure, we can find that every edge-label sequence is a
codeword of the code $(7,4)_{2}$, and vise versa. For example, the
following is a path in the trellis above:

{\tiny
$\left(\begin{array}{c}0\\0\\0\end{array}\right)\underline{1}\left(\begin{array}{c}1\\1\\0\end{array}\right)\underline{1}
\left(\begin{array}{c}0\\0\\1\end{array}\right)\underline{1}
\left(\begin{array}{c}0\\1\\0\end{array}\right)\underline{1}
\left(\begin{array}{c}0\\1\\1\end{array}\right)\underline{1}
\left(\begin{array}{c}1\\1\\1\end{array}\right)\underline{1}
\left(\begin{array}{c}1\\0\\1\end{array}\right)\underline{1}
\left(\begin{array}{c}0\\0\\0\end{array}\right)$ } The path above
represents a sequence $(1,1,1,1,1,1,1)$, which corresponds to the
codeword $[1,1, 1, 1, 1, 1, 1]$.

The trellis shown in Fig. 2 is a tail-biting trellis for the
$(7,4)_2$ Hamming code of Fig. 1.


\unitlength=0.11mm

\begin{picture}(500,820)(10,-560)

\scriptsize

\drawpath{43}{100}{100}{100}\put(80,100){$0$}
\put(96,100){$\begin{array}{c}|1|\\|1|\\|0|\end{array}$}

\drawpath{43}{-100}{100}{100}\put(80,80){$1$}
\put(96,100){$\begin{array}{c}|1|\\|1|\\|0|\end{array}$}

\drawpath{143}{100}{200}{100}\put(180,100){$0$}\put(196,100){$
\begin{array}{c}|1|\\|1|\\|0|\end{array}$}

\drawpath{243}{200}{300}{-200}\put(280,0){$0$}\put(296,-200){$
\begin{array}{c}|0|\\|0|\\|1|\end{array}$}

\drawpath{143}{100}{200}{200}\put(180,150){$1$}\put(196,200){$
\begin{array}{c}|0|\\|0|\\|1|\end{array}$}

\drawpath{243}{100}{300}{-400}\put(280,-390){$1$}\put(296,-400){$
\begin{array}{c}|1|\\|0|\\|1|\end{array}$}


\drawpath{43}{100}{100}{-100}\put(80,-100){$1$}\put(96,-100){$
\begin{array}{c}|0|\\|0|\\|0|\end{array}$}

\drawpath{43}{-100}{100}{-100}\put(80,-120){$0$}\put(96,-100){$
\begin{array}{c}|0|\\|0|\\|0|\end{array}$}

\drawpath{143}{-100}{200}{-100}\put(180,-100){$0$}\put(196,-100){$
\begin{array}{c}|0|\\|0|\\|0|\end{array}$}

\drawpath{243}{-100}{300}{-100}\put(280,-100){$0$}\put(296,-100){$
\begin{array}{c}|0|\\|0|\\|0|\end{array}$}

\drawpath{143}{-100}{200}{-200}\put(180,-150){$1$}\put(196,-200){$
\begin{array}{c}|1|\\|1|\\|1|\end{array}$}

\drawpath{243}{-200}{300}{-300}\put(260,-250){$1$}\put(296,-300){$
\begin{array}{c}|1|\\|0|\\|0|\end{array}$}



\drawpath{343}{-100}{400}{-100}\put(380,-100){$0$}\put(396,-100){$
\begin{array}{c}|0|\\|0|\\|0|\end{array}$}

\drawpath{343}{-200}{400}{-100}\put(380,-150){$1$}\put(396,-100){$
\begin{array}{c}|0|\\|0|\\|0|\end{array}$}

\drawpath{343}{-200}{400}{-200}\put(380,-200){$0$}\put(396,-200){$
\begin{array}{c}|0|\\|0|\\|1|\end{array}$}

\drawpath{343}{-100}{400}{-200}\put(380,-150){$1$}\put(396,-200){$
\begin{array}{c}|0|\\|0|\\|1|\end{array}$}

\drawpath{343}{-300}{400}{-300}\put(380,-300){$0$}\put(396,-300){$
\begin{array}{c}|1|\\|0|\\|0|\end{array}$}

\drawpath{343}{-400}{400}{-300}\put(380,-350){$1$}\put(396,-300){$
\begin{array}{c}|1|\\|0|\\|0|\end{array}$}

\drawpath{343}{-400}{400}{-400}\put(380,-400){$0$}\put(396,-400){$
\begin{array}{c}|1|\\|0|\\|1|\end{array}$}

\drawpath{343}{-300}{400}{-400}\put(380,-350){$1$}\put(396,-400){$
\begin{array}{c}|1|\\|0|\\|1|\end{array}$}


\drawpath{443}{-300}{500}{200}\put(480,180){$0$}\put(496,200){$
\begin{array}{c}|1|\\|0|\\|0|\end{array}$}

\drawpath{443}{-100}{500}{200}\put(480,200){$1$}\put(496,200){$
\begin{array}{c}|1|\\|0|\\|0|\end{array}$}

\drawpath{443}{-200}{500}{-200}\put(480,-200){$1$}\put(496,-200){$
\begin{array}{c}|1|\\|0|\\|1|\end{array}$}

\drawpath{443}{-400}{500}{-200}\put(480,-300){$0$}\put(496,-200){$
\begin{array}{c}|1|\\|0|\\|1|\end{array}$}

\drawpath{443}{-100}{500}{-100}\put(480,-100){$0$}\put(496,-100){$
\begin{array}{c}|0|\\|0|\\|0|\end{array}$}

\drawpath{443}{-300}{500}{-100}\put(480,-200){$1$}\put(496,-100){$
\begin{array}{c}|0|\\|0|\\|0|\end{array}$}

\drawpath{443}{-200}{500}{100}\put(496,60){$0$}\put(496,100){$
\begin{array}{c}|0|\\|0|\\|1|\end{array}$}

\drawpath{443}{-400}{500}{100}\put(496,40){$1$}\put(496,100){$
\begin{array}{c}|0|\\|0|\\|1|\end{array}$}


\drawpath{543}{-200}{600}{-200}\put(580,-200){$0$}\put(596,-200){$
\begin{array}{c}|1|\\|0|\\|1|\end{array}$}

\drawpath{543}{200}{600}{200}\put(580,200){$1$}\put(596,200){$
\begin{array}{c}|1|\\|1|\\|0|\end{array}$}

\drawpath{543}{-100}{600}{-100}\put(580,-100){$0$}\put(596,-100){$
\begin{array}{c}|0|\\|0|\\|0|\end{array}$}

\drawpath{543}{100}{600}{100}\put(580,100){$1$}\put(596,100){$
\begin{array}{c}|0|\\|1|\\|1|\end{array}$}


\drawpath{643}{200}{700}{100}\put(680,150){$0$}\put(696,100){$
\begin{array}{c}|1|\\|1|\\|0|\end{array}$}

\drawpath{643}{100}{700}{100}\put(680,100){$1$}\put(696,100){$
\begin{array}{c}|1|\\|1|\\|0|\end{array}$}

\drawpath{643}{-200}{700}{-100}\put(680,-150){$1$}\put(696,-100){$
\begin{array}{c}|0|\\|0|\\|0|\end{array}$}

\drawpath{643}{-100}{700}{-100}\put(680,-100){$0$}\put(696,-100){$
\begin{array}{c}|0|\\|0|\\|0|\end{array}$}


\put(0,100){$
\begin{array}{c}|1|\\|1|\\|0|\end{array}$}

\put(0,-100){$
\begin{array}{c}|0|\\|0|\\|0|\end{array}$}

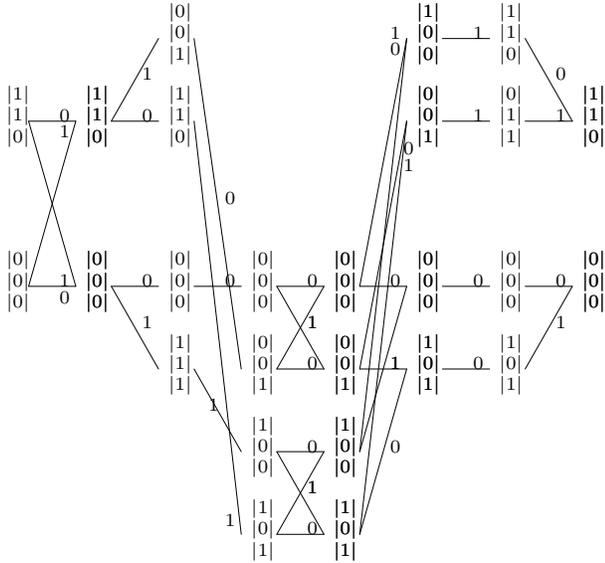
\put(60,-480){\small Fig.2 A tail-biting trellis for the $(7,4)_2$
 code in Fig.1.}


\end{picture}

Comparing to the figure 1, we can find that, in the figure 2, it has
two starting and ending pairs. Moreover, a cycle ( from the left
most vertex $ \left(\begin{array}{c}1\\1\\0\end{array}\right)$ to
the right most vertex
$\left(\begin{array}{c}1\\1\\0\end{array}\right)$ or from the left
most vertex $ \left(\begin{array}{c}0\\0\\0\end{array}\right)$ to
the right most vertex $
\left(\begin{array}{c}0\\0\\0\end{array}\right)$) corresponds a
codeword in figure 2, while a path corresponds a codeword in the
figure 1.

We also find that, in the figures, in addition to the labeling of
edges, each vertex in the set $V_i$ can be labeled by a sequence of
length $n-k$ of elements in $\Sigma$, and all vertex labels at a
given depth are distinct, just as shown in the figures 1 and 2.
Thus, every path (or cycle) in this labeled conventional trellis (
or tail-biting trellis) defines a sequence of length $n(1+n-k)$ over
$\Sigma$, consisting of alternating labels of vertices and edges in
$T$.  The set of all label sequences in a labeled trellis is
referred to as the label code represented by $T$ and is denoted by
$S(T)$. Fig.2 illustrates a labeled tail-biting trellis, and Fig.1
illustrates a labeled conventional trellis.

At last, we need two more definitions related to properties of a
trellis.

\noindent {\bf Definition 3}: A trellis $T$ is said to be linear
if there exists a vertex labeling of $T$ such that $S(T)$ is a
vector space.

The notion of mergeability \cite{Kschischang96,Vardy98,Vardy96} is
also useful here.

\noindent {\bf Definition 4}: A trellis is mergeable if there
exist vertices in the same vertex class of $T$ that can be
replaced by a single vertex, while retaining the edges incident on
the original vertices, without modifying $C(T)$. If a trellis
contains no vertices that can be merged, it is said to be
nonmergeable.

Koetter and Vardy \cite{Koetter02} have shown that if a linear
trellis is nonmergeable, then it is also biproper. However, though
the converse is true for conventional trellises, it is not true in
general for tail-biting trellises. They show that for tail-biting
trellises the following relation chain holds:

\begin{tabular}{c}
\{linear trellises\}\\
$\cup$\\
\{biproper linear trellises\}\\
$\cup$ \\
\{ nonmergeable linear trellises\}
\end{tabular}

In the discussion that follows, we restrict ourselves to trellises
representing linear block codes over the alphabet
$\Sigma=\mathbb{F}_q$. We will occasionally refer to vertices in a
trellis as ``states''.

\section{BCJR labeling and the embedding construction of
tail-biting trellis}

\subsection{The minimal BCJR labeling of a trellis}

The original BCJR algorithm [1] constructs the minimal and unique,
up to isomorphism, conventional trellis for a linear block code. In
the paper \cite{Nori}, the authors gave a simple method to describe
this construction. Here, we only give two examples to illustrate
this method. More details can be found in that paper.

%
%

\noindent {\bf Example 1}: Consider a self dual $(4,2)_2$ code with
parity check matrix defined as follows:
\begin{eqnarray*}
H= \left[
\begin{array}{cccc}
0 & 1&1&0\\
1 & 0&0&1
\end{array}\right].
\end{eqnarray*}
We obtain a minimal BCJR labeling of the trellis for the $(4,2)_2$
 code as illustrated in  Fig. 3.


\unitlength=0.11mm

\begin{picture}(300,550)(-100,-300)

\scriptsize

\drawpath{40}{00}{100}{100}\put(70,70){$1$}
\put(96,100){$\begin{array}{c}|0|\\|1|\end{array}$}

\drawpath{40}{00}{100}{-100}\put(70,-70){$0$}
\put(96,-100){$\begin{array}{c}|0|\\|0|\end{array}$}

\drawpath{140}{100}{200}{100}\put(170,100){$0$}
\put(196,100){$\begin{array}{c}|0|\\|1|\end{array}$}

\drawpath{140}{100}{200}{200}\put(170,170){$1$}
\put(196,200){$\begin{array}{c}|1|\\|1|\end{array}$}

\drawpath{140}{-100}{200}{-100}\put(170,-100){$0$}
\put(196,-100){$\begin{array}{c}|0|\\|0|\end{array}$}

\drawpath{140}{-100}{200}{-200}\put(170,-170){$1$}
\put(196,-200){$\begin{array}{c}|1|\\|0|\end{array}$}


\drawpath{240}{100}{300}{100}\put(270,100){$0$}
\put(296,100){$\begin{array}{c}|0|\\|1|\end{array}$}

\drawpath{240}{200}{300}{100}\put(270,170){$1$}
\put(296,100){$\begin{array}{c}|0|\\|1|\end{array}$}

\drawpath{240}{-100}{300}{-100}\put(270,-100){$0$}
\put(296,-100){$\begin{array}{c}|0|\\|0|\end{array}$}

\drawpath{240}{-200}{300}{-100}\put(270,-170){$1$}
\put(296,-100){$\begin{array}{c}|0|\\|0|\end{array}$}

\drawpath{340}{-100}{400}{00}\put(370,-70){$0$}
\put(396,00){$\begin{array}{c}|0|\\|0|\end{array}$}

\drawpath{340}{100}{400}{00}\put(370,70){$1$}
\put(396,00){$\begin{array}{c}|0|\\|0|\end{array}$}


\put(0,00){$
\begin{array}{c}|0|\\|0|\end{array}$}

\put(0,-260){\small  Fig.3 The minimal conventional trellis}

\put(90,-285){\small for the $(4,2)_2$ code.}
\end{picture}

\noindent {\bf Example 2}: Similarly, consider the $(7,4)_2$
Hamming code with parity check matrix defined as follows:
\begin{eqnarray*}
H= \left[
\begin{array}{ccccccc}
1 & 1&0&0&1&0&1\\
1 & 1&1&0&0&1&0\\
0 & 1&1&1&0&0&1
\end{array}\right].
\end{eqnarray*}
We obtain a minimal BCJR labeling of the trellis for the $(7,4)_2$
Hamming code as illustrated in  Fig. 1.

\subsection{The embedding construction of tail-biting trellis}

Now we can state our method to design a tail-biting trellis for a
given linear block code. This method is demonstrated by following
example.

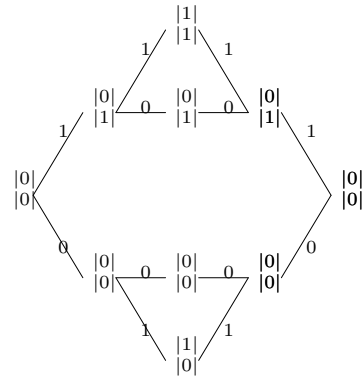
Let us first consider the minimal conventional trellis $T$ for the
$(4,2)_2$ code in Fig.3. Note that ${\bf
\alpha}=\left(\begin{array}{c}0\\1\end{array}\right)\in V_2$.
Arranging this vector $\alpha$ to the first column and the last
column in the parity check matrix $H$, we obtain
\begin{eqnarray*} H'= \left[
\begin{array}{cccccc}
0 & 0 & 1&1&0&0\\
1 & 1 & 0&0&1&1
\end{array}\right].
\end{eqnarray*}

Thus, we can get a minimal BCJR labeling of the trellis for the
parity check matrix $H'$ as illustrated in  Fig. 4.


\unitlength=0.11mm

\begin{picture}(300,570)(-100,-300)

\scriptsize

\drawpath{40}{100}{100}{100}\put(70,100){$0$}
\put(96,100){$\begin{array}{c}|0|\\|1|\end{array}$}

\drawpath{40}{-100}{100}{-100}\put(70,-100){$0$}
\put(96,-100){$\begin{array}{c}|0|\\|0|\end{array}$}

\drawpath{140}{100}{200}{100}\put(170,100){$0$}
\put(196,100){$\begin{array}{c}|0|\\|1|\end{array}$}

\drawpath{140}{100}{200}{200}\put(170,170){$1$}
\put(196,200){$\begin{array}{c}|1|\\|1|\end{array}$}

\drawpath{140}{-100}{200}{-100}\put(170,-100){$0$}
\put(196,-100){$\begin{array}{c}|0|\\|0|\end{array}$}

\drawpath{140}{-100}{200}{-200}\put(170,-170){$1$}
\put(196,-200){$\begin{array}{c}|1|\\|0|\end{array}$}


\drawpath{240}{100}{300}{100}\put(270,100){$0$}
\put(296,100){$\begin{array}{c}|0|\\|1|\end{array}$}

\drawpath{240}{200}{300}{100}\put(270,170){$1$}
\put(296,100){$\begin{array}{c}|0|\\|1|\end{array}$}

\drawpath{240}{-100}{300}{-100}\put(270,-100){$0$}
\put(296,-100){$\begin{array}{c}|0|\\|0|\end{array}$}

\drawpath{240}{-200}{300}{-100}\put(270,-170){$1$}
\put(296,-100){$\begin{array}{c}|0|\\|0|\end{array}$}



\drawpath{340}{100}{400}{100}\put(370,100){$0$}
\put(396,100){$\begin{array}{c}|0|\\|1|\end{array}$}

\drawpath{340}{100}{400}{-100}\put(370,00){$1$}
\put(396,-100){$\begin{array}{c}|0|\\|0|\end{array}$}

\drawpath{340}{-100}{400}{-100}\put(370,-100){$0$}
\put(396,-100){$\begin{array}{c}|0|\\|0|\end{array}$}

\drawpath{340}{-100}{400}{100}\put(370,00){$1$}
\put(396,100){$\begin{array}{c}|0|\\|1|\end{array}$}

\drawpath{440}{100}{500}{-00}\put(470,60){$1$}
\put(496,00){$\begin{array}{c}|0|\\|0|\end{array}$}

\drawpath{440}{-100}{500}{00}\put(470,-70){$0$}
\put(496,00){$\begin{array}{c}|0|\\|0|\end{array}$}

\drawpath{40}{100}{100}{-100}\put(70,0){$1$}
\put(96,-100){$\begin{array}{c}|0|\\|0|\end{array}$}

\drawpath{40}{-100}{100}{100}\put(70,-0){$1$}
\put(96,100){$\begin{array}{c}|0|\\|1|\end{array}$}

\drawpath{-60}{00}{00}{100}\put(-30,70){$1$} \put(0,100){$
\begin{array}{c}|0|\\|1|\end{array}$}

\put(-100,00){$
\begin{array}{c}|0|\\|0|\end{array}$}

\drawpath{-60}{00}{00}{-100}\put(-30,-80){$0$} \put(0,-100){$
\begin{array}{c}|0|\\|0|\end{array}$}

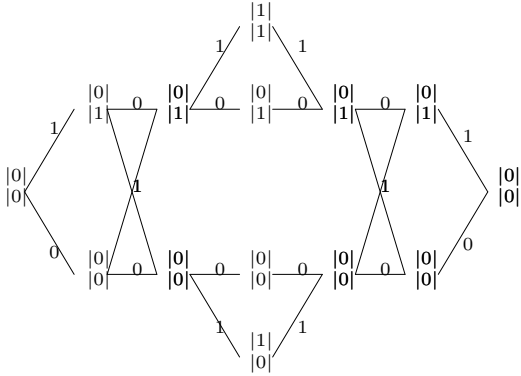
\put(0,-260){\small  Fig.4 The minimal conventional trellis}

\put(90,-285){\small for the parity check matrix $H'$.}
\end{picture}

In Fig.4, let $T_0$ be all paths from
$\left(\begin{array}{c}0\\0\end{array}\right)\in V'_1$ to
$\left(\begin{array}{c}0\\0\end{array}\right)\in V'_5$,  and $C_{0}$
be the set consisting of all codewords corresponding to $T_{0}$;
also let $T_1$ be all paths from
$\left(\begin{array}{c}0\\1\end{array}\right)\in V'_1$ to
$\left(\begin{array}{c}0\\1\end{array}\right)\in V'_5$, and $C_{1}$
be the set consisting of all codewords corresponding to $T_{1}$.
Comparing to Fig. 3, we can find out that both $C_0$ and $C_1$ are
the $(4,2)_2$ codewords.

Now let us consider the set $T_{0}\cap T_{1}$. This set can be
divided into two parts, moveover, these two parts are isomorphic. In
fact, the first part is consisted of the following vertexes:
$\left(\begin{array}{c}0\\1\end{array}\right)$,$\left(\begin{array}{c}0\\1\end{array}\right)$,$\left(\begin{array}{c}0\\1\end{array}\right)$
and $\left(\begin{array}{c}1\\1\end{array}\right)$, and the four
vertexes in the second part are
$\left(\begin{array}{c}0\\0\end{array}\right)$,$\left(\begin{array}{c}0\\0\end{array}\right)$,$\left(\begin{array}{c}0\\0\end{array}\right)$
and $\left(\begin{array}{c}1\\0\end{array}\right)$.

\par
Now we drop the four vertexes in the first part and the left most
and the right most vertexes from the figure 4 and obtain the
following trellis.
\unitlength=0.11mm

\begin{picture}(280,400)(460,-210)

\scriptsize

\put(496,100){$
\begin{array}{c}|0|\\|1|\end{array}$}

\put(496,-100){$
\begin{array}{c}|0|\\|0|\end{array}$}

\drawpath{540}{100}{600}{00}\put(570,55){$1$}
\put(596,00){$\begin{array}{c}|0|\\|0|\end{array}$}

\drawpath{540}{-100}{600}{00}\put(570,-70){$0$}
\put(596,00){$\begin{array}{c}|0|\\|0|\end{array}$}

\drawpath{640}{00}{700}{00}\put(670,0){$0$}
\put(696,00){$\begin{array}{c}|0|\\|0|\end{array}$}

\drawpath{640}{00}{700}{100}\put(670,70){$1$}
\put(696,100){$\begin{array}{c}|1|\\|0|\end{array}$}

\drawpath{740}{00}{800}{00}\put(770,0){$0$}
\put(796,00){$\begin{array}{c}|0|\\|0|\end{array}$}

\drawpath{740}{100}{800}{00}\put(770,65){$1$}
\put(796,00){$\begin{array}{c}|0|\\|0|\end{array}$}

\drawpath{840}{00}{900}{-100}\put(870,-75){$0$}
\put(896,-100){$\begin{array}{c}|0|\\|0|\end{array}$}

\drawpath{840}{00}{900}{100}\put(870,70){$1$}
\put(896,100){$\begin{array}{c}|0|\\|1|\end{array}$}

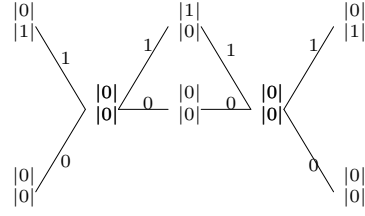
\put(490,-170){\small  Fig.5 The trellis constructed from Fig.4. }

\end{picture}

It is easy to verify that the codewords corresponding to Fig.5
compose the linear block code $(4,2)_{2}$. In fact, let us consider
the codewords in $C_0$ or $C_1$, passing only $V'_{3,0}=\{\left(
\begin{array}{c}0\\0\end{array}\right),\left(
\begin{array}{c}1\\0\end{array}\right)\}$.
Suppose $c\in C_1$, represented by the path $p$ not passing
$V'_{3,0}$. As $V'_{3,0}$ is a subspace of $V'_{3}$, the dimension
of $V'_{3,0}$ is one less than that of $V'_{3}$ and
$\left(\begin{array}{c}0\\1\end{array}\right)\notin V'_{3,0}$, thus
by adding $\left(\begin{array}{c}0\\1\end{array}\right)$ to each
vertex label in $p$, we get the path $p'$, passing $V'_{3,0}$. It is
clear that $p'$ represents a codeword $c\in C_0$. Similarly, suppose
$c\in C_0$, represented by a path  passing $V'_{3,0}$, then $c\in
C_1$, represented by a path not passing $V'_{3,0}$.
 Thus, the codewords passing only $V'_{3,0}$  in $C_0$ or
$C_1$ compose exactly the $(4,2)_2$ codewords.

Now we try to transform the operating steps above into a language of
parity check matrix. To get Fig. 5, we deleted half pathes in Fig.
4. In fact, it is equivalent to add a row to the parity check
matrix. Let us go to more detail.

Let $C'$ be the codewords with the parity check matrix $H'$, and
$C_t$  the codewords represented by all paths from
$\left(\begin{array}{c}0\\0\end{array}\right)\in V'_0$ to
$\left(\begin{array}{c}0\\0\end{array}\right)\in V'_6$, passing only
$V'_{3,0}$. As $V'_{3,0}$ is a subspace of $V'_{3}$ and the
dimension of $V'_{3,0}$ is one less than that of $V'_{3}$, thus the
dimension of $C_t$ is one less than that of $C'$. Therefore, there
exists a parity check matrix $H^{\dagger}$ for $C_t$, such that
$H^{\dagger}$ is obtained by adding  one more row to $H'$. In fact,
it is enough to let
\begin{eqnarray*} H^{\dagger}= \left[
\begin{array}{cccccc}
1 & 1 & 0&0&0&0\\
0 & 0 & 1&1&0&0\\
1 & 1 & 0&0&1&1
\end{array}\right].
\end{eqnarray*}
Furthermore, the minimal BCJR labeling of the trellis for the parity
check matrix $H^{\dagger}$ is illustrated in  Fig. 6.


\unitlength=0.11mm

\begin{picture}(300,550)(-100,-350)

\scriptsize

\drawpath{40}{-100}{100}{-100}\put(70,-100){$0$}
\put(96,-100){$\begin{array}{c}|0|\\|0|\\|0|\end{array}$}

\drawpath{140}{-100}{200}{-100}\put(170,-100){$0$}
\put(196,-100){$\begin{array}{c}|0|\\|0|\\|0|\end{array}$}

\drawpath{140}{-100}{200}{-200}\put(170,-170){$1$}
\put(196,-200){$\begin{array}{c}|0|\\|1|\\|0|\end{array}$}


\drawpath{240}{-100}{300}{-100}\put(270,-100){$0$}
\put(296,-100){$\begin{array}{c}|0|\\|0|\\|0|\end{array}$}

\drawpath{240}{-200}{300}{-100}\put(270,-170){$1$}
\put(296,-100){$\begin{array}{c}|0|\\|0|\\|0|\end{array}$}



\drawpath{340}{-100}{400}{-100}\put(370,-100){$0$}
\put(396,-100){$\begin{array}{c}|0|\\|0|\\|0|\end{array}$}

\drawpath{340}{-100}{400}{100}\put(370,00){$1$}
\put(396,100){$\begin{array}{c}|0|\\|0|\\|1|\end{array}$}

\drawpath{440}{100}{500}{-00}\put(470,60){$1$}
\put(496,00){$\begin{array}{c}|0|\\|0|\\|0|\end{array}$}

\drawpath{440}{-100}{500}{00}\put(470,-70){$0$}
\put(496,00){$\begin{array}{c}|0|\\|0|\\|0|\end{array}$}

\drawpath{40}{100}{100}{-100}\put(70,0){$1$}
\put(96,-100){$\begin{array}{c}|0|\\|0|\\|0|\end{array}$}

\drawpath{-60}{00}{00}{100}\put(-30,70){$1$} \put(0,100){$
\begin{array}{c}|1|\\|0|\\|1|\end{array}$}

\put(-100,00){$
\begin{array}{c}|0|\\|0|\\|0|\end{array}$}

\drawpath{-60}{00}{00}{-100}\put(-30,-80){$0$} \put(0,-100){$
\begin{array}{c}|0|\\|0|\\|0|\end{array}$}

\put(0,-270){\small  Fig.6 The minimal BCJR trellis for}

\put(90,-295){\small  the parity check matrix $H^\dagger$.}
\end{picture}

It is obvious that by deleting $V_0$ and $V_6$ and the corresponding
edges, and deleting the first row of each vertex label in Fig.6, we
get the Fig. 5, which turns out to be the labeled tail-biting
trellis for the $(4,2)_2$ code.

Now we generalize the operating steps shown in the above example
into a general operating method, and obtain the embedding
construction of a tail-biting trellis as follows:

\begin{description}
\item[1.] Let $C$ be an $(n,k)_q$ linear code with an  $(n-k)\times n$
parity check matrix
 $H =({\bf
h_1,h_2,\ldots,h_n})$,  and $T$ be its labeled BCJR trellis. Assume
that ${\bf\alpha}\in V_i ({\bf \alpha}\neq0)$.  Let $s_i$ denote the
dimension of $V_i,0\le i< n$. Since $V_i$ is a vector space, if
${\bf \alpha}\in V_i $, then $2{\bf\alpha}, 3{\bf\alpha}, \ldots,
(q-1){\bf\alpha}\in V_i$, there exists a linear subspace $V_{i,0}$
of dimension $s_i-1$, such that ${\bf \alpha}\notin V_{i,0}$. We now
add ${\bf \alpha}$ to  $H$ before the first column and after the
last column, respectively, and denote this new matrix as $H'$, that
is, $H' =({\bf \alpha, h_1,h_2,\ldots,h_n,\alpha})$. Construct a
labeled BCJR trellis $T'$ for $H'$.
\item[2.] Let $C_i$ be the codewords represented by all paths from
$i{\bf \alpha}\in V'_1$ to $i{\bf \alpha}\in V'_{n+1}, 0\le i\le
q-1$. Then $C_i$ is the  $(n,k)_q$ linear code $C$. Put
$V'_{i+1,0}=V_{i,0}$. Then $V'_{i+1,0}\subset V'_{i+1}$, and the
codewords only passing $V'_{i+1,0}$  in $C_0$ or $C_1$ or $\ldots$
or $C_{q-1}$ compose exactly the $(n,k)_q$ linear code $C$. Let
$C_t$ be the codewords represented by all paths passing only
$V'_{i+1,0}$. Compute the parity check matrix $H^{\dagger}$ for
$C_t$. Obviously, $H^{\dagger}$ has one more row than $H'$.
\item[3.] Let $T^{\dagger}$ be the labeled BCJR trellis for parity
check matrix $H^{\dagger}$. By deleting $V^{\dagger}_0$ and
$V^{\dagger}_{n+2}$ and relating edges, and deleting the first row
of each vertex label in $T^{\dagger}$, we get a labeled tail-biting
trellis for the $(n,k)_q$ linear code $C$.
\end{description}

It is easy to show the validity of the embedding construction. Thus,
with this new approach of constructing tail-biting trellises, most
of the study of tail-biting trellises can be converted into that of
conventional trellises.

Surprisingly, we can further process another embedding construction
based on the obtained labeled BCJR trellis $T^{\dagger}$.

For example, repeating the steps above on the parity check matrix
$H^{\dagger}$, which is corresponding to Fig. 6, we obtain a new
parity check matrix
\begin{eqnarray*} H^{\ddagger}= \left[
\begin{array}{cccccccc}
1 & 1 & 1 & 1&0&0&0&0\\
0 & 1 & 1 & 0&0&0&0&0\\
1 & 0 & 0 & 1&1&0&0&1\\
0 & 1 & 1 & 0&0&1&1&0
\end{array}\right].
\end{eqnarray*}
We can further get a BCJR trellis $T^{\ddagger}$ corresponding to
$H^{\ddagger}$, and we find that the dimension of $V^{\ddagger}_4$
is $0$.

Now some remarks are in order. The first one is that even if there
exists an integer $q'$ for ${\bf\alpha}\in V_i$, such that $0<q'<q$,
and $q'{\bf\alpha=0}$, ${\bf\alpha}, 2{\bf\alpha}, 3{\bf\alpha},
\ldots, (q-1){\bf\alpha}$ are not distinct, but the embedding
construction above can be similarly processed.

The second one is that ${\bf \alpha}\notin V_{i,0}$ is a necessary
condition. If ${\bf \alpha}\in V_{i,0}$, then the codewords passing
only $V'_{i+1,0}=V_{i,0}$  in $C_0$ or $C_1$ or $\ldots$ or
$C_{q-1}$ do not compose the $(n,k)_q$ linear code $C$.

The third one is that ${\bf\alpha}$, in fact, specifies a coset
decomposition $V_{i}/V_{i,0}$ of the vector space $V_{i}$, such that
every coset is associated with a unique $j{\bf\alpha}, 0\le j<q$.

The fourth one is to notice that $V_{i,0}$ is not necessarily
unique. For example, consider the trellis shown in Fig.3, let $i=2,
{\bf \alpha}=\left(\begin{array}{c}0\\1\end{array}\right)$. Then
 $V_{i,0}=\{\left(
\begin{array}{c}0\\0\end{array}\right),\left(
\begin{array}{c}1\\0\end{array}\right)\}$ or $\{\left(
\begin{array}{c}0\\0\end{array}\right),\left(
\begin{array}{c}1\\1\end{array}\right)\}$.
If $V_{i,0}=\{\left(
\begin{array}{c}0\\0\end{array}\right),\left(
\begin{array}{c}1\\1\end{array}\right)\}$, $H^{\dagger}$ will become

\begin{eqnarray*} H^{\dagger}= \left[
\begin{array}{cccccc}
1 & 1 & 1&0&0&0\\
0 & 0 & 1&1&0&0\\
1 & 1 & 0&0&1&1
\end{array}\right],
\end{eqnarray*}
and we will get another labeled tail-biting trellis for the
$(4,2)_2$ code as follows:


\unitlength=0.11mm

\begin{picture}(300,420)(-10,-240)

\scriptsize

\drawpath{40}{100}{100}{100}\put(70,100){$0$}
\put(96,100){$\begin{array}{c}|0|\\|1|\end{array}$}

\drawpath{40}{-100}{100}{-100}\put(70,-100){$0$}
\put(96,-100){$\begin{array}{c}|0|\\|0|\end{array}$}

\drawpath{140}{100}{200}{100}\put(170,100){$1$}
\put(196,100){$\begin{array}{c}|1|\\|1|\end{array}$}

\drawpath{140}{-100}{200}{-100}\put(170,-100){$0$}
\put(196,-100){$\begin{array}{c}|0|\\|0|\end{array}$}


\drawpath{240}{100}{300}{100}\put(270,100){$1$}
\put(296,100){$\begin{array}{c}|0|\\|1|\end{array}$}

\drawpath{240}{-100}{300}{-100}\put(270,-100){$0$}
\put(296,-100){$\begin{array}{c}|0|\\|0|\end{array}$}



\drawpath{340}{100}{400}{100}\put(370,100){$0$}
\put(396,100){$\begin{array}{c}|0|\\|1|\end{array}$}

\drawpath{340}{100}{400}{-100}\put(370,00){$1$}
\put(396,-100){$\begin{array}{c}|0|\\|0|\end{array}$}

\drawpath{340}{-100}{400}{-100}\put(370,-100){$0$}
\put(396,-100){$\begin{array}{c}|0|\\|0|\end{array}$}

\drawpath{340}{-100}{400}{100}\put(370,00){$1$}
\put(396,100){$\begin{array}{c}|0|\\|1|\end{array}$}


\drawpath{40}{100}{100}{-100}\put(70,0){$1$}
\put(96,-100){$\begin{array}{c}|0|\\|0|\end{array}$}

\drawpath{40}{-100}{100}{100}\put(70,-0){$1$}
\put(96,100){$\begin{array}{c}|0|\\|1|\end{array}$}

 \put(0,100){$
\begin{array}{c}|0|\\|1|\end{array}$}

 \put(0,-100){$
\begin{array}{c}|0|\\|0|\end{array}$}

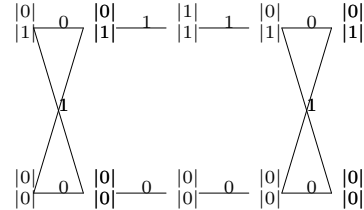
\put(-5,-190){\small  Fig.7 The trellis constructed from Fig.3.}

\end{picture}


The fifth one is that even though $V_{i,0}$ are different, the
corresponding tail-biting trellis is the same if $\alpha$ satisfies
some conditions. Give an example as follows:

 \noindent {\bf Example 3}:
Let $T$ be the labeled BCJR trellis in Fig.3. As ${\bf
\alpha}=\left(\begin{array}{c}0\\1\end{array}\right)\in V_1 \cap V_2
\cap V_3$, hence $V_{1,0}=\{\left(
\begin{array}{c}0\\0\end{array}\right)\}$, $V_{2,0}=\{\left(
\begin{array}{c}0\\0\end{array}\right),\left(
\begin{array}{c}1\\0\end{array}\right)\}$, $V_{3,0}=\{\left(
\begin{array}{c}0\\0\end{array}\right)\}$.
Obviously, the embedding construction by $V_{1,0}$ or $V_{2,0}$ or
$V_{3,0}$ gets the same tail-biting trellis.

To illustrate our method of construction, we demonstrate another
example.

\noindent {\bf Example 4}: Let $T$ be the trellis for the $(7,4)_2$
Hamming code in  Fig.1, and ${\bf \alpha}=\left(
\begin{array}{c}
1\\1\\0
\end{array}\right)$.
Similarly, the embedding construction by $V_{3,0}=\{\left(
\begin{array}{c}0\\0\\0\end{array}\right)$, $\left(
\begin{array}{c}0\\0\\1\end{array}\right), \left(
\begin{array}{c}1\\0\\0\end{array}\right), \left(
\begin{array}{c}1\\0\\1\end{array}\right)\}$ and
\begin{eqnarray*} H^{\dagger}= \left[
\begin{array}{ccccccccc}
1 & 1 & 1 & 1& 0&0&0&0&0\\
1 & 1 & 1 & 0 & 0 & 1&0&1&1\\
1 & 1 & 1 & 1 & 0 & 0 & 1&0&1\\
0 & 0 & 1 & 1 & 1 & 0 & 0 & 1&0
\end{array}\right],
\end{eqnarray*}
gets a labeled tail-biting trellis as illustrated in Fig. 2.\\

Furthermore, if we repeat the construction on the trellis
$T^{\dagger}$ with $H^{\dagger}$, a new tail-biting trellis can be
obtained as follows:  Take ${\bf \alpha}^{\dagger}=\left(
\begin{array}{c}
0\\1\\0\\1
\end{array}\right)$, $V^{\dagger}_{4,0}=\{\left(
\begin{array}{c}0\\0\\0\\0\end{array}\right)$, $\left(
\begin{array}{c}0\\0\\0\\1\end{array}\right)\}$ and get
\begin{eqnarray*} H^{\ddagger}= \left[
\begin{array}{ccccccccccc}
1 & 0 & 0 & 0 & 1 & 0 & 0 & 0 & 0 & 0 & 0\\
0 &1 & 1 & 1 & 1& 0&0&0&0&0&0\\
1 &1 & 1 & 1 & 0 & 0 & 1&0&1&1&1\\
0 &1 & 1 & 1 & 1 & 0 & 0 & 1&0&1&0\\
1 &0 & 0 & 1 & 1 & 1 & 0 & 0 & 1& 0&1
\end{array}\right],
\end{eqnarray*}
which generates another labeled tail-biting trellis, shown in the
following figure. Notice that, in this trellis, the dimensions of
both $V^{\ddagger}_5$ and $V^{\ddagger}_6$ are 1.


\unitlength=0.11mm

\begin{picture}(300,570)(10,-300)

\scriptsize

\drawpath{140}{100}{200}{100}\put(170,100){$1$}
\put(196,100){$\begin{array}{c}|0|\\|1|\\|0|\end{array}$}

\drawpath{140}{200}{200}{200}\put(170,200){$0$}
\put(196,200){$\begin{array}{c}|0|\\|1|\\|1|\end{array}$}

\drawpath{140}{-100}{200}{-100}\put(170,-100){$0$}
\put(196,-100){$\begin{array}{c}|0|\\|0|\\|0|\end{array}$}

\drawpath{140}{-200}{200}{-200}\put(170,-200){$1$}
\put(196,-200){$\begin{array}{c}|0|\\|0|\\|1|\end{array}$}


\drawpath{240}{100}{300}{100}\put(270,100){$1$}
\put(296,100){$\begin{array}{c}|0|\\|0|\\|1|\end{array}$}

\drawpath{240}{200}{300}{-100}\put(270,50){$1$}
\put(296,-100){$\begin{array}{c}|0|\\|0|\\|0|\end{array}$}

\drawpath{240}{-100}{300}{-100}\put(270,-100){$0$}
\put(296,-100){$\begin{array}{c}|0|\\|0|\\|0|\end{array}$}

\drawpath{240}{-200}{300}{100}\put(270,-50){$0$}
\put(296,100){$\begin{array}{c}|0|\\|0|\\|1|\end{array}$}



\drawpath{340}{100}{400}{100}\put(370,100){$0$}
\put(396,100){$\begin{array}{c}|0|\\|0|\\|1|\end{array}$}

\drawpath{340}{100}{400}{-100}\put(370,00){$1$}
\put(396,-100){$\begin{array}{c}|0|\\|0|\\|0|\end{array}$}

\drawpath{340}{-100}{400}{-100}\put(370,-100){$0$}
\put(396,-100){$\begin{array}{c}|0|\\|0|\\|0|\end{array}$}

\drawpath{340}{-100}{400}{100}\put(370,00){$1$}
\put(396,100){$\begin{array}{c}|0|\\|0|\\|1|\end{array}$}


\drawpath{440}{100}{500}{200}\put(470,170){$0$}
\put(496,200){$\begin{array}{c}|0|\\|0|\\|1|\end{array}$}

\drawpath{440}{100}{500}{100}\put(470,100){$1$}
\put(496,100){$\begin{array}{c}|1|\\|0|\\|1|\end{array}$}

\drawpath{440}{-100}{500}{-100}\put(470,-100){$0$}
\put(496,-100){$\begin{array}{c}|0|\\|0|\\|0|\end{array}$}

\drawpath{440}{-100}{500}{-200}\put(470,-150){$1$}
\put(496,-200){$\begin{array}{c}|1|\\|0|\\|0|\end{array}$}

\drawpath{540}{200}{600}{200}\put(570,200){$1$}
\put(596,200){$\begin{array}{c}|0|\\|1|\\|1|\end{array}$}

\drawpath{540}{100}{600}{100}\put(570,100){$0$}
\put(596,100){$\begin{array}{c}|1|\\|0|\\|1|\end{array}$}

\drawpath{540}{-200}{600}{-200}\put(570,-200){$1$}
\put(596,-200){$\begin{array}{c}|1|\\|1|\\|0|\end{array}$}

\drawpath{540}{-100}{600}{-100}\put(570,-100){$0$}
\put(596,-100){$\begin{array}{c}|0|\\|0|\\|0|\end{array}$}


\drawpath{640}{200}{700}{200}\put(670,200){$0$}
\put(696,200){$\begin{array}{c}|0|\\|1|\\|1|\end{array}$}

\drawpath{640}{100}{700}{-100}\put(670,0){$1$}
\put(696,-100){$\begin{array}{c}|0|\\|0|\\|0|\end{array}$}

\drawpath{640}{100}{700}{100}\put(670,100){$0$}
\put(696,100){$\begin{array}{c}|1|\\|0|\\|1|\end{array}$}

\drawpath{640}{200}{700}{-200}
\put(696,-200){$\begin{array}{c}|1|\\|1|\\|0|\end{array}$}

\drawpath{640}{-100}{700}{100}\put(670,0){$1$}
\put(696,100){$\begin{array}{c}|1|\\|0|\\|1|\end{array}$}

\drawpath{640}{-100}{700}{-100}\put(670,-100){$0$}
\put(696,-100){$\begin{array}{c}|0|\\|0|\\|0|\end{array}$}

\drawpath{640}{-200}{700}{200}
\put(696,200){$\begin{array}{c}|0|\\|1|\\|1|\end{array}$}

\drawpath{640}{-200}{700}{-200}\put(670,-200){$0$}
\put(696,-200){$\begin{array}{c}|1|\\|1|\\|0|\end{array}$}

\drawpath{40}{200}{100}{200}\put(70,200){$0$}
\put(96,200){$\begin{array}{c}|0|\\|1|\\|1|\end{array}$}

\drawpath{40}{200}{100}{100}\put(70,170){$1$}
\put(96,100){$\begin{array}{c}|1|\\|0|\\|1|\end{array}$}

\drawpath{40}{100}{100}{100}\put(70,100){$0$}
\put(96,100){$\begin{array}{c}|1|\\|0|\\|1|\end{array}$}

\drawpath{40}{100}{100}{200}\put(70,170){$1$}
\put(96,200){$\begin{array}{c}|0|\\|1|\\|1|\end{array}$}

 \put(0,200){$
\begin{array}{c}|0|\\|1|\\|1|\end{array}$}

 \put(0,100){$
\begin{array}{c}|1|\\|0|\\|1|\end{array}$}

 \put(0,-100){$
\begin{array}{c}|0|\\|0|\\|0|\end{array}$}

 \put(0,-200){$
\begin{array}{c}|1|\\|1|\\|0|\end{array}$}

\drawpath{40}{-100}{100}{-100}\put(70,-100){$0$}
\put(96,-100){$\begin{array}{c}|0|\\|0|\\|0|\end{array}$}

\drawpath{40}{-100}{100}{-200}\put(70,-180){$1$}
\put(96,-200){$\begin{array}{c}|1|\\|1|\\|0|\end{array}$}

\drawpath{40}{-200}{100}{-100}\put(70,-180){$1$}
\put(96,-100){$\begin{array}{c}|0|\\|0|\\|0|\end{array}$}

\drawpath{40}{-200}{100}{-200}\put(70,-200){$0$}
\put(96,-200){$\begin{array}{c}|1|\\|1|\\|0|\end{array}$}

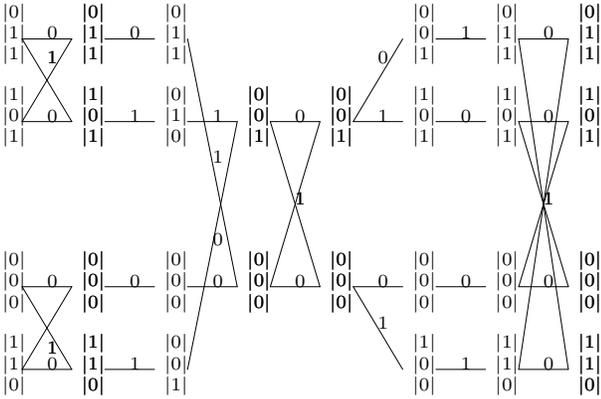
\put(35,-280){\small  Fig.8 The tail-biting trellis for the parity
check matrix $H^\ddagger$.}

\end{picture}

\subsection{Results on the embedding construction}

For our construction above, some properties are important.

\noindent {\bf Lemma 1}: Let $T$ be a trellis for an $(n,k)_q$
linear code $C$ with the parity check matrix $H =({\bf
h_1,h_2,\ldots,h_n})$. Suppose ${\bf\alpha}\in V_i ({\bf
\alpha}\neq0)$, $V_{i,0}$ be a linear subspace of $V_i$ of dimension
$s_i-1$, such that ${\bf \alpha}\notin V_{i,0}$. Let $H' =({\bf
\alpha, h_1,h_2,\ldots,h_n,\alpha})$, and $T'$ a labeled BCJR
trellis for $H'$. Let $C_t$ be the codewords represented by all
paths passing only $V'_{i+1,0}$. Suppose $H^{\dagger}$ is an
embedding construction by ${\bf\alpha}$ and $V_{i,0}$, and
$H^{\dagger}$ has
 one more row $(x_1,x_2,\ldots,x_{n+2})$ than $H'$. Then,

\begin{enumerate}
\item $(x_1,x_2,\ldots,x_{n+2})$ is not unique;

\item $x_1$ and $x_{n+2}$ are distinct;

\item $(x_1,x_2,\ldots,x_{n+2})$ can be
$(1,x_2,\dots,x_{i+1},0,\ldots,0)$, such that for each
$(c_1,c_2,\ldots,c_{n+2})\in C_t$,
$c_1+x_2c_2+x_3c_3+\ldots+x_{i+1}c_{i+1}=0$.
\end{enumerate}

\begin{proof}
1) As parity check matrix $H^{\dagger}$ is not unique, so does
$(x_1,x_2,\ldots,x_{n+2})$;

2) Note that $(1,0,0,\ldots,0,q-1)$ represents a path in $T'$ and
$(1,0,0,\ldots,0,q-1)\notin C_t$, it is clear that $x_1$ and
$x_{n+2}$ are distinct;

3) Suppose that there is a row vector
$(1,x_2,\dots,x_{i+1},0,\ldots,0)$, such that for each
$(c_1,c_2,\ldots,c_{n+2})\in C_t$,
$c_1+x_2c_2+x_3c_3+\ldots+x_{i+1}c_{i+1}=0$, and $H^{\dagger}$ has
 one more row $(1,x_2,\dots,x_{i+1},0,\ldots,0)$ than $H'$. As $x_1=1$ and
$x_{n+2}=0$ are distinct, hence the rank of $H^{\dagger}$ is one
more than that of $H'$, thus $H^{\dagger}$ is the parity check
matrix for $C_t$.

Now we show the existence of the row vector
$(1,x_2,\dots,x_{i+1},0,\ldots,0)$.

Let $C(T')$ denote the code represented by the trellis $T'$. Let
$C'_{i+1}=\{(c_1,c_2,\ldots,c_{i+1})|(c_1,c_2,\ldots,c_{n+2})\in
C(T')\}$,
$C_{i+1,t}=\{(c_1,c_2,\ldots,c_{i+1})|(c_1,c_2,\ldots,c_{n+2})\in
C_t\}$.

As $C(T')$ and $C_t$ are linear code, so do $C'_{i+1}$ and
$C_{i+1,t}$, and $C_{i+1,t}$ is a true linear subspace of
$C'_{i+1}$. Thus, there is a row vector $(x_1,x_2,\dots,x_{i+1})$,
such that for each $(c_1,c_2,\ldots,c_{i+1})\in C_{i+1,t}$,
$x_1c_1+x_2c_2+x_3c_3+\ldots+x_{i+1}c_{i+1}=0$, and for each
$(c_1,c_2,\ldots,c_{i+1})\in C'_{i+1}\setminus C_{i+1,t}$,
$x_1c_1+x_2c_2+x_3c_3+\ldots+x_{i+1}c_{i+1}\ne 0$. Note that
$(1,0,\ldots,0)\in C'_{i+1}\setminus C_{i+1,t}$, so $x_1\ne 0$.
\end{proof}

\noindent {\bf Theorem 1}: The tail-biting  trellis  for an linear
block code $(n,k)_q$, got by an embedding construction, is linear
and non-mergeable.

\begin{proof} It is well known that the labeled BCJR trellis for an linear
block code  is  nonmergeable and linear. $T^\dag$ is the labeled
BCJR trellis for parity check matrix $H^{\dagger}$, and the
tail-biting  trellis is got from $T^\dag$, hence linear and
non-mergeable.
\end{proof}

We call ${\bf b}\in V_{i+1}$ is the map of ${\bf a}\in V_{i}$,
denoted by $M(a)$, if there exits an edge from ${\bf a}$ to ${\bf
b}$. Note that $M(a)$ is not necessarily unique. Further, let
$M^1(a)$=$M(a)$, $M^r(a)$ the map of  $M^{r-1}(a)$, where $r>1$.

From Theorem 1, we prove the following lemma.

 \noindent {\bf Lemma 2}: Let $V^\dag_i, 0<i<n$, denotes the state
space of the trellis $T^\dag$ got by an embedding construction with
${\bf \alpha}$ and $V_{i,0}$ from trellis $T$.
 Let $M(V_{i}), M(V_{i,0})$ denote the map of $V_{i}, V_{i,0}$,
respectively. Then  $M(V_{i,0})$ is a vector space. And,

 Case 1.
$M^r(V_{i})=M^r(V_{i,0})$ and ${\bf \alpha}\in M^r(V_{i,0})$. Then
$V^\dag_{i+r}=M^r(V_{i})$.

 Case 2.
$M^r(V_{i})=M^r(V_{i,0})$ and ${\bf \alpha}\notin M^r(V_{i,0})$.
Then $V^\dag_{i+r}$ is a vector space generated by $M^r(V_{i})$ and
${\bf \alpha}$.

 Case 3.
$M^r(V_{i})\neq M^r(V_{i,0})$ and all $M^r({\bf \alpha})-{\bf
\alpha}\in M^r(V_{i,0})$. Then $V^\dag_{i+r}=M^r(V_{i,0})$.

 Case 4.
$M^r(V_{i})\neq M^r(V_{i,0})$ and  not all $M^r({\bf \alpha})-{\bf
\alpha}\in M^r(V_{i,0})$. Then $V^\dag_{i+r}$ is a vector space
generated by $M^r(V_{i,0})$ and $M^r({\bf \alpha})-{\bf \alpha}$,
here we select $M^r({\bf \alpha})$ such that $M^r({\bf \alpha})-{\bf
\alpha}\notin M^r(V_{i,0})$.

\begin{proof}
From Theorem 1, it is known that $V^\dag_i$ is a vector space.
 We now show that $M(V_{i,0})$ is a vector space.

Let $a,b\in M(V_{i,0})$. Then there exist $x,y\in S(T)$, such that
$x_i,y_i\in V_{i,0}$, and $a=x_{i+1}, b=y_{i+1}$, here $z_i$ denotes
a state label of $z\in S(T)$ at time index $i$. From $x+y\in S(T)$
and $x_i+y_i\in V_{i,0}$, we have $a+b=x_{i+1}+y_{i+1}\in
M(V_{i,0})$, hence $M(V_{i,0})$ is a vector space and so is
$M^r(V_{i,0})$ for $r>1$.

We only prove the Case 4. The others are similar.

 Case 1.
$M^r(V_{i})=M^r(V_{i,0})$ and ${\bf \alpha}\in M^r(V_{i,0})$.  The
trellis in Fig.8 for $i=4$ and $r=1$ or 2 belongs to this case.

 Case 2.
$M^r(V_{i})=M^r(V_{i,0})$ and ${\bf \alpha}\notin M^r(V_{i,0})$.
 The trellis in Fig.2 for $i=4$ and $r=2$ belongs to this
case.

 Case 3.
$M^r(V_{i})\neq M^r(V_{i,0})$ and all $M^r({\bf \alpha})-{\bf
\alpha}\in M^r(V_{i,0})$.  The trellis in Fig.2 for $i=3$  and $r=1$
belongs to this case.

 Case 4.
$M^r(V_{i})\neq M^r(V_{i,0})$ and  not all $M^r({\bf \alpha})-{\bf
\alpha}\in M^r(V_{i,0})$.  The trellis in Fig.2 for $i=4$ and $r=1$
 belongs to this case.

Note that $(q-j){\bf \alpha}+M^r(j{\bf \alpha}+{\bf \beta})\in
V^\dag_{i+r}$, $0\le j<q$, $\beta \in V_{i,0}$. For any state $j{\bf
\alpha}+{\bf \beta}\in V_{i}$,
 we know that
\begin{eqnarray*}
M^r(j{\bf \alpha}+{\bf \beta})&=& M^r(j{\bf \alpha})+M^r({\bf
\beta})\\
&=& jM^r({\bf \alpha})+M^r({\bf \beta})\\
\therefore (q-j){\bf \alpha}+M^r(j{\bf \alpha}+{\bf \beta})&=&
j(M^r({\bf \alpha})-{\bf \alpha})+M^r({\bf \beta})
\end{eqnarray*}

This completes the proof.
\end{proof}

In a similar way to the above discussion, one may discuss the case
for $0<j<i$.\\

\noindent {\bf Lemma 3}: Let $T$ be a trellis for an $(n,k)_q$
linear code $C$. Suppose ${\bf\alpha}\in V_i ({\bf \alpha}\neq0)$,
$V_{i,0}$ be a linear subspace of $V_i$ of dimension $s_i-1$, such
that ${\bf \alpha}\notin V_{i,0}$.  Then we can get a tail-biting
trellis $T^\dagger$ with an embedding construction by ${\bf\alpha}$
and $V_{i,0}$, such that the dimension of $V^\dagger_i$ is $s_i-1$.

\begin{proof}  Let $H =({\bf
h_1,h_2,\ldots,h_n})$ be a parity check matrix for $T$, and let $H'
=({\bf \alpha, h_1,h_2,\ldots,h_n,\alpha})$. Construct a labeled
conventional trellis $T'$ for $H'$.

Let $C_i$ be the codewords represented by all paths from $i{\bf
\alpha}\in V'_1$ to $i{\bf \alpha}\in V'_{n+1}, 0\le i\le q-1$. Then
$C_i$ is the  linear code for $T$.

Note that all paths from ${\bf 0}\in V'_1$ to ${\bf 0}\in V'_{n+1}$
compose exactly the trellis $T$, and adding $i{\bf \alpha}$ to each
vertex label in all paths from ${\bf 0}\in V'_1$ to ${\bf 0}\in
V'_{n+1}$ compose exactly all paths from $i{\bf \alpha}\in V'_1$ to
$i{\bf \alpha}\in V'_{n+1}, 0< i\le q-1$. As $i{\bf \alpha}\in
V_{i}$, thus $V'_{i+1}=V_{i}$.

By the process of embedding construction with ${\bf\alpha}$ and
$V_{i,0}$,  it is clear that we can get a tail-biting trellis
$T^\dagger$, such that the dimension of $V^\dagger_i$ is $s_i-1$.
\end{proof}

An embedding construction has two key parameters: ${\bf \alpha}$ and
 $V'_{i,0}$. Therefore, to construct a minimal
tail-biting trellis is to determine the sequence of ${\bf \alpha}$
and $V'_{i,0}$.

Now we can state one of the main results as a theorem.

 \noindent {\bf Theorem 2}:  Any minimal
tail-biting trellis for an $(n,k)_q$ linear code can be constructed
by embedding constructions from a Bahl-Cocke-Jelinek-Raviv(BCJR)
constructed conventional trellis.

\begin{proof}
Let $T$ be a minimal tail-biting trellis. Suppose $\alpha\in V_0$
but $\alpha\notin V_i$. From $T$, construct a new tail-biting $T'$
starting at time index $i$, i.e. $V'_0=V_i, \ldots, V'_{n-i}=V_0,
V'_{n-i+1}=V_1, \ldots, V'_{n-1}=V_{i-1}$.

From Lemma 3, the dimension of $V'_{n-i}$ can be reduced by 1, i.e.
the dimension of $V_{0}$ can be reduced by 1.

Repeat the process above, we get a tail-biting trellis $T^\dag$,
such that $V^\dag_0=\{{\bf 0}\}$. As the
Bahl-Cocke-Jelinek-Raviv(BCJR) constructed conventional trellis is
unique, we know that $T^\dag$ is a BCJR constructed conventional
trellis.

Therefore, to construct a minimal tail-biting trellis, one just need
to process conversely from $T^\dag$.
\end{proof}

\section{To reduce the maximum state-complexity of a tail-biting trellis with one peak}

In this section, we restrict ourselves to trellises representing
binary linear block codes.

Using embedding constructions, we discuss how to reduce the maximum
state-complexity of a tail-biting (or conventional) trellis with one
peak.

We first consider the following simplest case.

\noindent {\bf Proposition 1}: Let $T$ be a  trellis. Suppose
$|V_p|>|V_{p-1}|$ and $|V_p|>|V_{p+1}|$, where $1<p<n-1$, and
$|V_{p-1}|\ge4$.  we also assume that $|V_i|<|V_{p-1}|$ for $0\le
i<p-1$ and $p+1< i<n$. Then the maximum state-complexity of  $T$ can
be reduced by 1 with an embedding construction.

\begin{proof}

We first show that  $ |V_{p-1}\cap V_{p}\cap V_{p+1}|>1$.

Suppose  $V_{p-1}=\{\alpha_0,\alpha_1,\ldots,\alpha_{k-1} \}$.
Then
$V_{p}=\{\alpha_0,\alpha_1,\ldots,\alpha_{k-1},\alpha_0+\beta,\alpha_1+\beta,\ldots,\alpha_{k-1}+\beta
\}$, and $V_{p+1}\subset V_{p}, |V_{p+1}|=|V_{p-1}|$.

From $|V_{p-1}|\ge4$, it is easy to see that there exist
$\alpha_i,\alpha_j\in V_{p+1}$ or
$\alpha_i+\beta,\alpha_j+\beta\in V_{p+1}$, where
$\alpha_i\neq\alpha_j$.

If $\alpha_i,\alpha_j\in V_{p+1}$, then assume $\alpha_j\neq
\mathbf{0}$,  hence $ |V_{p-1}\cap V_{p}\cap V_{p+1}|>1$.

If $\alpha_i+\beta,\alpha_j+\beta\in V_{p+1}$, then
$\alpha_i+\beta+\alpha_j+\beta=\alpha_i+\alpha_j \neq \mathbf{0}$,
and $\alpha_i+\alpha_j\in V_{p+1}$, hence $ |V_{p-1}\cap V_{p}\cap
V_{p+1}|>1$.

 Let
$\alpha\in V_{p-1}\cap V_{p}\cap V_{p+1}, \alpha\neq \mathbf{0}$.
Let $s_i$ denote the dimension of $V_i,0\le i< n$. A linear subspace
$V_{p,0}$ of  dimension $s_p-1$ is existed, such that $
V_{p,0}\subset V_{p}$ and $\alpha\notin V_{p,0}$.

Let $T^\dag$ be the trellis got by an embedding construction with
$\alpha$ and $V_{p,0}$. It is easy to show that the maximum
state-complexity of $T^\dag$ is one less than that of $T$.

\end{proof}

To prove the following proposition, we first state a lemma.

\noindent {\bf Lemma 4}: Let $T$ be a  trellis. For
$i\in\{0,1,\ldots,n-1\}$, every vertex of $V_i$ has the same  out
degree 1 or 2.

\begin{proof}By the definition of the trellis for a linear code, every vertex of $V_i$ has at least  out
degree 1. If we note the following  fact, then the proof is obvious.

For $\alpha\in V_i, \alpha\neq \mathbf{0}$, the out degree of
$\mathbf{0}$ is 2$\Longleftrightarrow$  there exists a codeword
${\bf c}=(0,\ldots,0,1,c_{i+2},\ldots,c_n)$ $\Longleftrightarrow$
the out degree of $\alpha$ is 2.

\end{proof}

\noindent {\bf Proposition 2}: Let $T$ be a  trellis. Suppose
$|V_p|>|V_{p-1}|$, $|V_p|=|V_{p+1}|$ and $|V_{p+1}|>|V_{p+2}|$,
where  $1<p<n-2$, and $|V_{p-1}|\ge 8$. We also assume that
$|V_i|<|V_{p-1}|$ for $0\le i<p-1$ and $p+2< i<n$. Then the maximum
state-complexity of  $T$ can be reduced by 1 with an embedding
construction.

\begin{proof}
Let ${\bf h_1,h_2,\ldots,h_n}$ be the $n$ columns of $H$.

First consider the case that ${\bf h_{p+1}}\in V_p$. Then
$V_p=V_{p+1}$. Now we show that $ |V_{p-1}\cap V_{p}\cap V_{p+1}\cap
V_{p+2}|>3$.

Suppose  $V_{p-1}=\{\alpha_0,\alpha_1,\ldots,\alpha_{k-1} \}$.
Then

$V_{p}=\{\alpha_0,\alpha_1,\ldots,\alpha_{k-1},\alpha_0+\beta,\alpha_1+\beta,\ldots,\alpha_{k-1}+\beta
\}$,

$ V_{p+2}\subset V_{p}, |V_{p+2}|=|V_{p-1}|$.

From $|V_{p-1}|\ge8$, it is easy to see that there exist
$\alpha_i,\alpha_j, \alpha_r, \alpha_s\in V_{p+2}$, or
$\alpha_i+\beta,\alpha_j+\beta, \alpha_r+\beta, \alpha_s+\beta\in
V_{p+2}$, where $\alpha_i, \alpha_j, \alpha_r, \alpha_s$  are
distinct.

If $\alpha_i,\alpha_j, \alpha_r, \alpha_s\in V_{p+2}$, then $
|V_{p-1}\cap V_{p}\cap V_{p+1}\cap V_{p+2}|>3$.

If $\alpha_i+\beta,\alpha_j+\beta,\alpha_r+\beta, \alpha_s+\beta\in
V_{p+2}$, then  $\alpha_i+\alpha_j,
\alpha_i+\alpha_r,\alpha_i+\alpha_s\in V_{p+2}$, hence $
|V_{p-1}\cap V_{p}\cap V_{p+1}\cap V_{p+2}|>3$.

Suppose that $\alpha, \beta\in V_{p-1}\cap V_{p}\cap V_{p+1}\cap
V_{p+2}, \alpha\neq\beta, \alpha\neq \mathbf{0}, \beta\neq
\mathbf{0}$.

Let $\alpha={\bf h_{p+1}}$. Then a linear subspace $V_{p,0}$ of
dimension $s_p-1$ is existed, such that $ V_{p,0}\subset V_{p}$ and
$\alpha\in V_{p,0}, \beta\notin V_{p,0}$.

 From the proof of Lemma 2,
we know that $M(V_{p,0})$ is also a vector space, where $M(V_{p,0})$
denotes the map of $V_{p,0}$. As $\alpha\in V_{p,0}, \beta\notin
V_{p,0}$, thus $M(V_{p,0})\subseteq V_{p,0}$, hence $\beta\notin
M(V_{p,0})$.

If $M(\mathbf{0})=\{\mathbf{0}\}$, then $|M(V_{p,0})|=|V_{p,0}|$,
hence $M(V_{p,0})= V_{p,0}$.

If $M(\mathbf{0})=\{\mathbf{0},\alpha\}$, then $V_{p,0}\subseteq
M(V_{p,0})$, hence $M(V_{p,0})= V_{p,0}$.

Let $T^\dag$ be the trellis got by an embedding construction with
$\beta$ and $V_{p,0}$. It is easy to show that the maximum
state-complexity of $T^\dag$ is one less than that of $T$.

Let $\alpha\neq{\bf h_{p+1}}$. Then a linear subspace $V_{p,0}$ of
dimension $s_p-1$ is existed, such that $ V_{p,0}\subset V_{p}$ and
$\alpha\notin V_{p,0}, {\bf h_{p+1}}\in V_{p,0}$. Similarly, we know
that  $M(V_{p,0})= V_{p,0}$, and $\alpha\notin M(V_{p,0})$.

Let $T^\dag$ be the trellis got by an embedding construction with
$\alpha$ and $V_{p,0}$. It is easy to show that the maximum
state-complexity of $T^\dag$ is one less than that of $T$.

Second consider the case that ${\bf h_{p+1}}\notin V_p$. Then
 the out degree of  every vertex in $V_p$ is 1 as $|V_p|=|V_{p+1}|$.
Now we show that $ |V_{p-1}\cap V_{p}\cap V_{p+1}\cap V_{p+2}|>1$.

Suppose  $V_{p-1}=\{\alpha_0,\alpha_1,\ldots,\alpha_{k-1} \}$. Then

$V_{p}=\{\alpha_0,\alpha_1,\ldots,\alpha_{k-1},\alpha_0+\beta,\alpha_1+\beta,\ldots,\alpha_{k-1}+\beta
\}$,

$V_{p+1}\subset\{\alpha_0,\alpha_1,\ldots,\alpha_{k-1},\alpha_0+\beta,\alpha_1+\beta,\ldots,\alpha_{k-1}+\beta
,\\
\alpha_0+\gamma,\alpha_1+\gamma,\ldots,\alpha_{k-1}+\gamma,\alpha_0+\beta+\gamma,\alpha_1+\beta+\gamma,\ldots,\alpha_{k-1}+\beta+\gamma
\}$, and $ V_{p+2}\subset V_{p+1}, |V_{p+2}|=|V_{p-1}|$.

From $|V_{p-1}|\ge8$, it is easy to see that there exist
$\alpha_i,\alpha_j\in V_{p+2}$, or $\alpha_i+\beta,\alpha_j+\beta\in
V_{p+2}$, or $\alpha_i+\gamma,\alpha_j+\gamma\in V_{p+2}$ or
$\alpha_i+\beta+\gamma,\alpha_j+\beta+\gamma\in V_{p+2}$, where
$\alpha_i\neq\alpha_j$.

If $\alpha_i,\alpha_j\in V_{p+2}$, then assume $\alpha_j\neq
\mathbf{0}$,  hence $ |V_{p-1}\cap V_{p}\cap V_{p+1}\cap
V_{p+2}|>1$.

If $\alpha_i+\beta,\alpha_j+\beta\in V_{p+2}$, then
$\alpha_i+\beta+\alpha_j+\beta=\alpha_i+\alpha_j \neq \mathbf{0}$,
and $\alpha_i+\alpha_j\in V_{p+2}$, hence $ |V_{p-1}\cap V_{p}\cap
V_{p+1}\cap V_{p+2}|>1$.

  The other cases are similar.

Suppose that $\alpha\in V_{p-1}\cap V_{p}\cap V_{p+1}\cap V_{p+2},
 \alpha\neq \mathbf{0}$.

We first show that $M(\alpha)=\alpha$.

Suppose that $M(\alpha)=\alpha+{\bf h_{p+1}}$ and
$M(\gamma)=\gamma+{\bf h_{p+1}}=\alpha$. Then $\gamma={\bf
h_{p+1}}+\alpha$, which implies that ${\bf h_{p+1}}=\gamma+\alpha\in
V_p$. This is a contradiction.

Then a linear subspace $V_{p,0}$ of dimension $s_p-1$ is existed,
such that $ V_{p,0}\subset V_{p}$ and $\alpha\notin V_{p,0}$. Then
both $V_{p,0}$ and $M(V_{p,0})$ has the dimension  $s_p-1$, and
$\alpha\notin M(V_{p,0})$.
\end{proof}

Now we consider the  trellis $T$ illustrated in Fig.1. Let
$\alpha=\left(
\begin{array}{c}1\\1\\1\end{array}\right)$. Then $\alpha\in V_{2}\cap V_{3}\cap V_{4}\cap
V_{5}$.

Let $V_{3,0}=\{\left(
\begin{array}{c}0\\0\\0\end{array}\right)$, $\left(
\begin{array}{c}0\\0\\1\end{array}\right)$, $\left(
\begin{array}{c}1\\0\\0\end{array}\right)$, $\left(
\begin{array}{c}1\\0\\1\end{array}\right)\}$ in $T$. Then
$\alpha\notin M(V_{3,0})$.

With an embedding construction by $ \alpha$, $V_{3,0}$, and
\begin{eqnarray*} H^{\dagger}= \left[
\begin{array}{ccccccccc}
1 & 1 & 1 & 1 & 0 & 0 & 0 & 0&0\\
1 & 1 & 1 & 0 & 0 & 1 & 0 & 1&1\\
1 & 1 & 1 & 1 & 0 & 0 & 1 & 0&1\\
1 & 0 & 1 & 1 & 1 & 0 & 0 & 1&1
\end{array}\right],
\end{eqnarray*} we obtain
the  trellis in Fig.9.


\unitlength=0.11mm

\begin{picture}(500,900)(10,-560)

\scriptsize

\drawpath{43}{-100}{100}{-100}\put(80,-100){$0$}
\put(96,-100){$\begin{array}{c}|0|\\|0|\\|0|\end{array}$}

\drawpath{43}{-100}{100}{100}\put(80,80){$1$}
\put(96,100){$\begin{array}{c}|1|\\|1|\\|0|\end{array}$}

\drawpath{143}{100}{200}{100}\put(180,100){$0$}\put(196,100){$
\begin{array}{c}|1|\\|1|\\|0|\end{array}$}

\drawpath{143}{200}{200}{100}\put(180,150){$1$}\put(196,100){$
\begin{array}{c}|1|\\|1|\\|0|\end{array}$}

\drawpath{243}{200}{300}{-200}\put(280,0){$0$}\put(296,-200){$
\begin{array}{c}|0|\\|0|\\|1|\end{array}$}

\drawpath{143}{200}{200}{200}\put(180,200){$0$}\put(196,200){$
\begin{array}{c}|0|\\|0|\\|1|\end{array}$}

\drawpath{143}{100}{200}{200}\put(180,150){$1$}\put(196,200){$
\begin{array}{c}|0|\\|0|\\|1|\end{array}$}

\drawpath{243}{100}{300}{-400}\put(280,-390){$1$}\put(296,-400){$
\begin{array}{c}|1|\\|0|\\|1|\end{array}$}


\drawpath{43}{100}{100}{-200}\put(80,-200){$0$}\put(96,-200){$
\begin{array}{c}|1|\\|1|\\|1|\end{array}$}

\drawpath{43}{100}{100}{200}\put(80,200){$1$}\put(96,200){$
\begin{array}{c}|0|\\|0|\\|1|\end{array}$}

\drawpath{143}{-100}{200}{-100}\put(180,-100){$0$}\put(196,-100){$
\begin{array}{c}|0|\\|0|\\|0|\end{array}$}

\drawpath{143}{-200}{200}{-100}\put(180,-150){$1$}\put(196,-100){$
\begin{array}{c}|0|\\|0|\\|0|\end{array}$}

\drawpath{243}{-100}{300}{-100}\put(280,-100){$0$}\put(296,-100){$
\begin{array}{c}|0|\\|0|\\|0|\end{array}$}

\drawpath{143}{-200}{200}{-200}\put(180,-200){$0$}\put(196,-200){$
\begin{array}{c}|1|\\|1|\\|1|\end{array}$}

\drawpath{143}{-100}{200}{-200}\put(180,-150){$1$}\put(196,-200){$
\begin{array}{c}|1|\\|1|\\|1|\end{array}$}

\drawpath{243}{-200}{300}{-300}\put(260,-250){$1$}\put(296,-300){$
\begin{array}{c}|1|\\|0|\\|0|\end{array}$}



\drawpath{343}{-100}{400}{-100}\put(380,-100){$0$}\put(396,-100){$
\begin{array}{c}|0|\\|0|\\|0|\end{array}$}

\drawpath{343}{-200}{400}{-100}\put(380,-150){$1$}\put(396,-100){$
\begin{array}{c}|0|\\|0|\\|0|\end{array}$}

\drawpath{343}{-200}{400}{-200}\put(380,-200){$0$}\put(396,-200){$
\begin{array}{c}|0|\\|0|\\|1|\end{array}$}

\drawpath{343}{-100}{400}{-200}\put(380,-150){$1$}\put(396,-200){$
\begin{array}{c}|0|\\|0|\\|1|\end{array}$}

\drawpath{343}{-300}{400}{-300}\put(380,-300){$0$}\put(396,-300){$
\begin{array}{c}|1|\\|0|\\|0|\end{array}$}

\drawpath{343}{-400}{400}{-300}\put(380,-350){$1$}\put(396,-300){$
\begin{array}{c}|1|\\|0|\\|0|\end{array}$}

\drawpath{343}{-400}{400}{-400}\put(380,-400){$0$}\put(396,-400){$
\begin{array}{c}|1|\\|0|\\|1|\end{array}$}

\drawpath{343}{-300}{400}{-400}\put(380,-350){$1$}\put(396,-400){$
\begin{array}{c}|1|\\|0|\\|1|\end{array}$}


\drawpath{443}{-200}{500}{-200}\put(480,-200){$1$}\put(496,-200){$
\begin{array}{c}|1|\\|0|\\|1|\end{array}$}

\drawpath{443}{-400}{500}{-200}\put(480,-300){$0$}\put(496,-200){$
\begin{array}{c}|1|\\|0|\\|1|\end{array}$}

\drawpath{443}{-100}{500}{-100}\put(480,-100){$0$}\put(496,-100){$
\begin{array}{c}|0|\\|0|\\|0|\end{array}$}

\drawpath{443}{-300}{500}{-100}\put(480,-200){$1$}\put(496,-100){$
\begin{array}{c}|0|\\|0|\\|0|\end{array}$}


\drawpath{543}{-200}{600}{-200}\put(580,-200){$0$}\put(596,-200){$
\begin{array}{c}|1|\\|0|\\|1|\end{array}$}

\drawpath{543}{-100}{600}{200}\put(580,200){$1$}\put(596,200){$
\begin{array}{c}|0|\\|1|\\|0|\end{array}$}

\drawpath{543}{-100}{600}{-100}\put(580,-100){$0$}\put(596,-100){$
\begin{array}{c}|0|\\|0|\\|0|\end{array}$}

\drawpath{543}{-200}{600}{100}\put(580,100){$1$}\put(596,100){$
\begin{array}{c}|1|\\|1|\\|1|\end{array}$}


\drawpath{643}{200}{700}{100}\put(680,150){$1$}\put(696,100){$
\begin{array}{c}|1|\\|1|\\|1|\end{array}$}

\drawpath{643}{100}{700}{100}\put(680,100){$0$}\put(696,100){$
\begin{array}{c}|1|\\|1|\\|1|\end{array}$}

\drawpath{643}{-200}{700}{-100}\put(680,-150){$1$}\put(696,-100){$
\begin{array}{c}|0|\\|0|\\|0|\end{array}$}

\drawpath{643}{-100}{700}{-100}\put(680,-100){$0$}\put(696,-100){$
\begin{array}{c}|0|\\|0|\\|0|\end{array}$}


\put(0,100){$
\begin{array}{c}|1|\\|1|\\|1|\end{array}$}

\put(0,-100){$
\begin{array}{c}|0|\\|0|\\|0|\end{array}$}

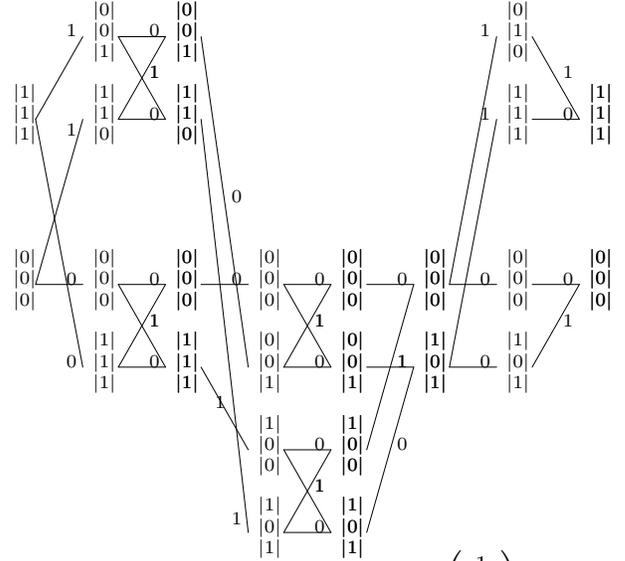
\put(30,-480){\small Fig.9 An embedding construction by ${\bf
\alpha}=\left(
\begin{array}{c}
1\\1\\1
\end{array}\right)$ and $V_{3,0}$.}

\end{picture}

With a similar argument as Proposition 2, we have the following
proposition.

\noindent {\bf Proposition 3}:
 Let $T$ be a  trellis. Suppose
$|V_p|>|V_{p-1}|$, $V_p=V_{p+1}=V_{p+2}$ and $|V_{p+2}|>|V_{p+3}|$,
where  $1<p<n-3$, and $|V_{p-1}|\ge 8$. We also assume that
$|V_i|<|V_{p-1}|$ for $0\le i<p-1$ and $p+3< i<n$. Then the maximum
state-complexity of  $T$ can be reduced by 1 with an embedding
construction.

\noindent {\bf Proposition 4}:
 Let $T$ be a  trellis. Suppose
$|V_p|>|V_{p-1}|$, $|V_p|=|V_{p+1}|=|V_{p+2}|, V_p\neq V_{p+1}$ or $
V_{p+1}\neq V_{p+2}$ and $|V_{p+2}|>|V_{p+3}|$, where $1<p<n-3$, and
$|V_{p-1}|\ge 16$. We also assume that $|V_i|<|V_{p-1}|$ for $0\le
i<p-1$ and $p+3< i<n$. Then the maximum state-complexity of  $T$ can
be reduced by 1 with an embedding construction.

\begin{proof} We just show the case that $V_p\neq V_{p+1} =
V_{p+2}$. The others are similar.

With a similar argument as Proposition 2, we may show that $
|V_{p-1}\cap V_{p}\cap V_{p+1}\cap V_{p+2}\cap V_{p+3}|>3$.

Suppose that $\alpha, \beta\in V_{p-1}\cap V_{p}\cap V_{p+1}\cap
V_{p+2}\cap V_{p+3}, \alpha\neq\beta, \alpha\neq \mathbf{0},
\beta\neq \mathbf{0}$.

If $\beta={\bf h_{p+2}}$. Note that for  $\alpha, \beta\in V_{p},
M(\alpha)=\alpha, M(\beta)=\beta$. Then a linear subspace $V_{p,0}$
of $V_{p}$ of dimension $s_p-1$ is existed, such that $\alpha\notin
V_{p,0}$. Then both $V_{p,0}$ and $M(V_{p,0})$ has the dimension
$s_p-1$, and $\alpha\notin M(V_{p,0})$, $\beta\in M(V_{p,0})$. Hence
$M^2(V_{p,0})=M(V_{p,0})$, and $\alpha\notin M^2(V_{p,0})$. With an
embedding construction by $\alpha$ and $V_{p,0}$, we have the
proposition.

If $\alpha\neq{\bf h_{p+2}}$ and  $\beta\neq{\bf h_{p+2}}$. Then
$V_{p+2}$ has a linear subspace $V_{p+2,0}$  of dimension $s_p-1$,
such that  $\alpha\notin V_{p+2,0}, {\bf h_{p+2}}\in V_{p+2,0}$.
Then  $M^{-1}(V_{p+2,0})=V_{p+2,0}$, where $M^{-1}(V_{p+2,0})$
denotes the set $U\subset V_{p+1} $, such that $M(U)=V_{p+2,0}$.
Hence $\alpha\notin M^{-2}(V_{p+2,0})$, $M^{-2}(V_{p+2,0})\subset
V_{p}, M^{-2}(V_{p+2,0})$ has the dimension $s_p-1$. With an
embedding construction by $\alpha$ and $M^{-2}(V_{p+2,0})$, we have
the proposition.

\end{proof}

Similarly, we may further discuss how to reduce the maximum
state-complexity of the trellis with one peak and
$|V_p|=|V_{p+1}|=\cdots=|V_{p+j}|$ for $j>2$.

\section{Conclusion}
We have presented a new approach of constructing tail-biting
trellises for linear block codes, and have proved that any minimal
tail-biting trellis can be constructed by the recursive process of
embedding constructions from a BCJR constructed conventional
trellis. We conclude this paper by observing that the minimal
tail-biting trellis computation problem may thus be stated as
follows:

Find the least embedding constructions, such that the minimal
tail-biting trellis can be constructed  from a BCJR constructed
conventional trellis.

 \section*{ Acknowledgment}
The authors would like to thank Dr. Haiquan Wang for a great
improvement of the writing of this paper. The research was supported
by Zhejiang Natural Science Foundation (No.Y1100318, R1090138) and Chinese Natural Science Foundation
(No. 60802047).


{\bf Jianqin Zhou} received his B.Sc. degree in mathematics from
East China Normal University, China, in 1983, and M.Sc. degree in
probability and statistics from Fudan University, China, in 1989.
From 1989 to 1999 he was with the Department of Mathematics and
Computer Science, Qufu Normal University, China. From 2000 to 2002,
he worked for a number of IT companies in Japan. From 2003 to 2007
he was with the Department of Computer Science, Anhui University of
Technology, China.
 From Sep 2006  to
Feb 2007, he was a visiting scholar with the Department of
Information and Computer Science, Keio University, Japan. Since 2008
he has been with the Telecommunication School, Hangzhou Dianzi
University, China

He  published more than  70 papers, and  proved a conjecture posed
by famous mathematician Paul Erd\H{o}s et al. His research interests
include coding theory, cryptography and combinatorics.

\end{document}